\documentclass[twocolumn]{aastex63}

\usepackage{amsmath}

\received{2020 October 5}
\revised{2020 November 13}
\accepted{2020 November 14}
\shorttitle{A type IV burst from Proxima Cen}
\shortauthors{Zic et al.}
\graphicspath{{./}{}}

\begin{document}

\title{A Flare--Type IV Burst Event from Proxima Centauri and Implications for Space Weather}
\correspondingauthor{Andrew Zic}
\email{azic7771@uni.sydney.edu.au}

\author[0000-0002-9583-2947]{Andrew Zic}
\affiliation{Sydney Institute for Astronomy, School of Physics, University of Sydney, NSW 2006, Australia}
\affiliation{CSIRO Astronomy and Space Science, PO Box 76, Epping, NSW 1710, Australia}

\author[0000-0002-2686-438X]{Tara Murphy}
\affiliation{Sydney Institute for Astronomy, School of Physics, University of Sydney, NSW 2006, Australia}
\affiliation{ARC Centre of Excellence for Gravitational Wave Discovery (OzGrav), Hawthorn 3122, VIC, Australia}

\author[0000-0002-0494-192X]{Christene Lynch}
\affiliation{International Centre for Radio Astronomy Research (ICRAR), Curtin University, Bentley, WA, Australia}
\affiliation{ARC Centre of Excellence for All Sky Astrophysics in 3 Dimensions (ASTRO3D), Bentley, WA, Australia}

\author[0000-0002-2155-6054]{George Heald}
\affiliation{CSIRO Astronomy and Space Science, P.O. Box 1130, Bentley, WA 6102, Australia} 

\author[0000-0002-9994-1593]{Emil Lenc}
\affiliation{CSIRO Astronomy and Space Science, P.O. Box 76, Epping, NSW 1710, Australia}

\author[0000-0001-6295-2881]{David L. Kaplan}
\affiliation{Department of Physics, University of Wisconsin Milwaukee, Milwaukee, Wisconsin 53201, USA}

\author[0000-0001-6978-9765]{Iver H. Cairns}
\affiliation{School of Physics, University of Sydney, NSW 2006, Australia}

\author{David Coward}
\affiliation{OzGrav-UWA, University of Western Australia, Department of Physics, M013, 35 Stirling Highway, Crawley, WA 6009, Australia} 

\author[0000-0002-9077-2025]{Bruce Gendre}
\affiliation{OzGrav-UWA, University of Western Australia, Department of Physics, M013, 35 Stirling Highway, Crawley, WA 6009, Australia} 

\author[0000-0002-6169-614X]{Helen Johnston}
\affiliation{Sydney Institute for Astronomy, School of Physics, University of Sydney, NSW 2006, Australia}

\author[0000-0001-7891-8143]{Meredith MacGregor}
\affiliation{Department of Astrophysical and Planetary Sciences, University of Colorado, 2000 Colorado Avenue, Boulder, CO 80309, USA}

\author[0000-0003-2783-1608]{Danny C. Price}
\affiliation{Department of Astronomy, University of California Berkeley, Berkeley CA 94720, USA}
\affiliation{Centre for Astrophysics \& Supercomputing, Swinburne University of Technology, Hawthorn, VIC 3122, Australia}

\author[0000-0001-5100-2354]{Michael S. Wheatland}
\affiliation{Sydney Institute for Astronomy, School of Physics, University of Sydney, NSW 2006, Australia}



\begin{abstract}
Studies of solar radio bursts play an important role in understanding the dynamics and acceleration processes behind solar space weather events, and the influence of solar magnetic activity on solar system planets. Similar low-frequency bursts detected from active M-dwarfs are expected to probe their space weather environments and therefore the habitability of their planetary companions.
Active M-dwarfs produce frequent, powerful flares which, along with radio emission, reveal conditions within their atmospheres. However, to date, only one candidate solar-like coherent radio burst has been identified from these stars, preventing robust observational constraints on their space weather environment.
During simultaneous optical and radio monitoring of the nearby dM5.5e star Proxima Centauri, we detected a bright, long-duration optical flare, accompanied by a series of intense, coherent radio bursts. These detections include the first example of an interferometrically detected coherent stellar radio burst temporally coincident with a flare, strongly indicating a causal relationship between these transient events. 
The polarization and temporal structure of the trailing long-duration burst enable us to identify it as a type IV burst. This represents the most compelling detection of a solar-like radio burst from another star to date. Solar type IV bursts are strongly associated with space weather events such as coronal mass ejections and solar energetic particle events, suggesting that stellar type IV bursts may be used as a tracer of stellar coronal mass ejections. We discuss the implications of this event for the occurrence of coronal mass ejections from Proxima Cen and other active M-dwarfs.
\end{abstract}

%

\section{Introduction}
    M-dwarfs are the most populous type of star in the Galaxy \citep{2006AJ....132.2360H} and have a high rate of close-in terrestrial exoplanets, $\sim 1.2$ per star \citep{2019AJ....158...75H}. Active M-dwarfs frequently produce flares several orders of magnitude more energetic than solar flares \citep{1976ApJS...30...85L}. Associated increases in ionizing radiation, along with frequent impacts of space weather events, such as coronal mass ejections (CMEs), are expected to result in magnetospheric compression and atmospheric erosion of close-in planetary companions, threatening their habitability \citep{2007AsBio...7..167K, 2007AsBio...7..185L}. 
    While the radiative components of M-dwarf flares are routinely detected and characterized across all wavelengths, their phenomenology arises from processes confined to the stellar atmosphere. As a result, the space weather environment around these stars cannot be assessed from flare observations alone. {Nonetheless, there have been promising developments in observational signatures of stellar CMEs in recent years. For example, \citet{2019NatAs...3..742A} reported the first confirmed stellar CME, detected via X-ray spectroscopy of a flare on the giant star HR~9024. Candidate CME events from active M-dwarfs have been inferred through signatures such as prolonged X-ray absorption \citep{2019ApJ...877..105M, 2017ApJ...850..191M}, and blue-shifted Balmer line components \citep{2019A&A...623A..49V,2011A&A...536A..62L,1990A&A...238..249H}.} However, these may be solely flare-related phenomena, such as chromospheric evaporation or as a result of limited diagnostic ability of low-resolution X-ray spectroscopy \citep{2019NatAs...3..742A, 2017IAUS..328..243O}. 
    
    Another promising route toward characterizing space weather around active M-dwarfs is the detection of solar-like type II, III, and IV bursts \citep{1950AuSRA...3..387W, 1957CRAS..244.1326B, white07, 2017IAUS..328..243O, 
    2020A&A...639L...7V}. These low-frequency radio bursts ($\lesssim 1$\,GHz) trace distinct particle acceleration processes within and beyond the corona that contribute to space weather, and are classified according to their morphologies in dynamic spectra \citep[e.g.][]{1963ARA&A...1..291W, white07}.
    
    \subsection{Characteristics of Space Weather-Related Radio Bursts}
    \label{spaceweather_bursts}

    Solar type II, type III, and type IV bursts have the following characteristics and interpretations: type II bursts last several minutes and exhibit a relatively slow drift from high to low frequencies, and are produced by CME shock fronts or flare blast waves propagating outward through the corona -- {see \citet{2003SSRv..107...27C} and \citet{2011sswh.book..267C} for a review of type II bursts in the context of solar system shocks}. There have been several efforts toward detecting stellar type II bursts, with no positive detections \citep{2016ApJ...830...24C,2018ApJ...856...39C,2018ApJ...862..113C,2019ApJ...871..214V}. Type III bursts are short-lived ($\sim 0.1\,$s), rapidly drifting bursts driven by relativistic electron beams escaping from the corona along open magnetic field lines -- {see \citep{2014RAA....14..773R} for a review}. Type IV bursts are long-duration bursts that occur during and after the decay phase of large flares \citep{1963PASJ...15..327T}, and are thought to be driven by continuous injection of energetic electrons into post-flare magnetic structures following CMEs \citep{2011ApJ...743..145C,2020A&A...639A.102S}.  We do not expand upon solar type I and type V bursts here as they are not relevant to space weather studies \citep{white07}, {but refer the reader to texts such as \citet{1965sra..book.....K} and 
    \citet{1963ARA&A...1..291W} for overviews and detailed definitions of solar radio bursts and other emissions}.
    
    Type IV bursts are particularly important in solar space weather studies, because they are associated with space weather events such as CMEs and solar energetic particle (SEP) events, and because they indicate ongoing electron acceleration following large flares \citep[e.g.][]{1992JGR....97.1619K}. 
    For example, \citet{1986SoPh..104...33R} argued that CMEs are a \textit{necessary} condition for the generation  of type IV bursts, and \citet{1988ApJ...325..895C} also argued that type IV bursts occur with CMEs, which also explains their association with SEPs \citep{1982ApJ...261..710K}. \citet{1988ApJ...325..901C} found that 88\% of type IV bursts are associated with type II bursts, with almost all interplanetary type II bursts associated with CMEs \citep{2003SSRv..107...27C}. In addition, solar type II and type IV bursts are strongly associated with H$\alpha$ flares \citep{1988ApJ...325..901C}. Type IV bursts 
    show a wide range of features, and consist of several subclasses \citep{1986SoPh..104...19P}. Of particular interest here are the ``decimetric type IV bursts'' (type IVdm) \citep{1976ApJ...204..597B}, 
    which have a frequency range spanning $\sim200$--$2000$\,MHz \citep{2011ApJ...743..145C}, covering the frequency range accessible with current low and mid-frequency radio facilities. These bursts have long delays ($> 30$\,minutes) from the microwave continuum emission peak, 
    and are composed of several subcomponents lasting tens of minutes to several hours \citep{1986SoPh..104...19P}. The early component is often complex, showing a variety of frequency drifts \citep{1967SoPh....1..304T}. In addition, they often exhibit high degrees of circular polarization (up to 100\%), 
    fractional bandwidths $\Delta \nu / \nu \sim 0.1$--$1$ \citep{1976ApJ...204..597B, 2011ApJ...743..145C}, and intensities reaching up to $10^{6}$\,sfu ($10^{10}$\,Jy), making them among the most intense solar radio bursts recorded \citep{2011ApJ...743..145C}. The associated brightness temperatures are very high (between $10^{8}$\,K and $10^{15}$\,K){, with brightness temperatures in excess of $10^{12}\,\text{K}$ requiring a coherent mechanism \citep{1969ApJ...155L..71K}. At the lower range of brightness temperatures ($\lesssim 10^{11}\,\text{K}$) additional properties such as narrow spectral features and high degrees of circular polarization may indicate a coherent emission mechanism, though incoherent gyrosynchrotron emission is regularly invoked for bursts at this lower range of brightness temperatures lacking these indicators of coherence \citep[][]{2019A&A...623A..63M}}. If the bursts exhibit a frequency drift, it is usually small (drift rates $|\dot{\nu}| \leq 100$\,kHz\,s$^{-1}$), and may indicate a gradually-expanding source region with decreasing magnetic field strength and/or plasma density \citep{1963PASJ...15..327T}.
  
    \subsection{Stellar Radio Bursts in the Solar Paradigm}
    The detection of solar-like radio bursts holds great potential for understanding coronal particle acceleration processes in M-dwarf flares, and for diagnosing the space weather environment around these stars. For example, solar type IV bursts are associated with space weather events such as CMEs and SEP events \citep{1986SoPh..104...33R,1988ApJ...325..901C, 2020A&A...639A.102S}, and probe ongoing electron acceleration in magnetic structures following large flares \citep{2020A&A...639A.102S,2011ApJ...743..145C, 1963ARA&A...1..291W}. 
    
    {
    However, there have been few unambiguous identifications of solar-like low-frequency bursts from M-dwarfs or other stellar systems to date. \citet{1982ApJ...252..239K} detected a strong radio burst from the dM4.0e star YZ Canis Minoris in time-series data with the Jodrell Bank interferometer at 408\,MHz, beginning $\sim 17$\,minutes after flaring activity in optical and X-ray wave bands. Owing to the delayed onset of the burst, \citet{1982ApJ...252..239K} identified this radio event as a type IV burst. Other early low-frequency detections were made with time-series data from single-dish telescopes, making them susceptible to terrestrial interference \citep{1990SoPh..130..265B, 1990ApJ...353..265B}. For example, \citet{1976ApJ...203..497S} and \citet{1969Natur.222.1126L} detected several M-dwarf radio bursts with intensities ranging from hundreds of millijansky to several jansky, coincident with optical flaring activity. However, low-frequency interferometric observations of active M-dwarfs have struggled to detect bursts of similar intensities or at similar rates as recorded in early single-dish observations \citep{2019ApJ...871..214V, 2017ApJ...836L..30L, 1978Natur.273..644D}, casting doubt on the reliability of these early detections. A recent exception to this trend of faint, low-duty cycle bursts at low radio frequencies is the detection of a 5.9\,Jy burst from dM4e star AD Leonis at 73.5\,MHz reported by \citet{2020MNRAS.494.4848D}. Although the signal-to-noise ratio of this detection is fairly low, future interferometric detections at these very low frequencies ($< 100$\,MHz) may provide some validation for early single-dish detections.}
    
    Coherent radio bursts from M-dwarfs detected with modern interferometric facilities have shown properties at odds with solar observations. Foremost, the majority of observations show poor association between radio bursts and multiwavelength flaring activity \citep{2018ApJ...856...39C, 1990SoPh..130..265B, 1988A&A...195..159K, 1981ApJ...245.1009H}, with the exception of the results from \citet{1982ApJ...252..239K} described above. Another key contrast is that there have been no morphological classifications of solar-like radio bursts to date, although some stellar radio bursts have shown spectro-temporal features consistent with solar radio bursts --- e.g., ``sudden reductions'' or  ``quasiperiodic pulsations'' \citep{1990ApJ...353..265B}. A final point of difference is that coherent radio bursts from M-dwarfs consistently exhibit high degrees of circular polarization ($f_C \sim 50$--$100\%$; \citep{2019ApJ...871..214V}), whereas only some solar radio bursts, such as type I, type IV, and decimetric spike bursts, are highly polarized \citep{1962PASJ...14....1K, 1965PASJ...17..294K, 1986SoPh..104...57A} -- other solar radio bursts exhibit only mild degrees of polarization. 
    
    These differences have hampered efforts to understand M-dwarf radio activity based on our more complete understanding of the Sun \citep{2019ApJ...871..214V}. 
    {In addition, recent studies have shown that the low-frequency variability of active \citep{2019MNRAS.488..559Z} and inactive \citep{2020NatAs...4..577V} M-dwarfs may arise from auroral processes in their magnetospheres. These phenomena are driven by ongoing field-aligned currents in the strong, large-scale magnetic field of the star, rather than flaring activity associated with localized active regions. This suggests that in general, the physical driver of many low-frequency radio bursts from M-dwarfs may be decoupled from the flares probed by optical and X-ray wave bands---in stark contrast to the Sun.}
    
    \subsection{Outline of This Paper}
     In this article, we report the detection of several radio bursts from Proxima Centauri (hereafter Proxima Cen) associated with a large optical flare. In Section \ref{obs} we detail the multiwavelength observations and data reduction. In Section \ref{multiwave_results} we describe the detections of the flare and radio bursts. We discuss these detections in light of the solar paradigm, and possible implications for space weather around Proxima Cen. In Section \ref{proxima_concl} we summarize our findings.

\section{Multiwavelength Observations and Data Reduction}
\label{obs}
    To search for space weather signatures from an active M-dwarf, we observed Proxima Cen simultaneously at multiple wavelengths over 11 nights.
    This star is suitable for our study because it is close to the Sun \citep[1.3\,pc;][]{2018A&A...616A...1G}, is magnetically active, and hosts a terrestrial-size planet within its habitable zone \citep{2016Natur.536..437A}, along with a recently discovered planet candidate at 1.5\,AU \citep{2020SciA....6.7467D}. Proxima Cen is also an interesting target because its slower rotation and slightly more mild activity levels \citep{2007AcA....57..149K, 2008A&A...489L..45R} differentiate it from more active and rapidly rotating M-dwarfs that have been previously targeted in search of space weather events \citep{2018ApJ...856...39C, 2018ApJ...862..113C, 2016ApJ...830...24C, 2019ApJ...871..214V}.
    
    {Before providing details on observations and data reduction for the facilities used in this work, we note that all local observatory times have been converted to the barycentric dynamical timeframe (TDB), to ensure consistency between ground- and space-based facilities.}
    \subsection{Transiting Exoplanet Survey Satellite (TESS)}
        NASA's TESS \citep{2015JATIS...1a4003R} observed Proxima Cen during the period of 2019 April--May in its Sector 11 observing run. Observations were taken through the broad red bandpass spanning $5813$--$11159$\,\AA\ on the TESS instrument (the TESS band). We downloaded the calibrated TESS light curves for Proxima Centauri from the Mikulski Archive for Space Telescopes\footnote{\url{https://archive.stsci.edu/}}. Light curves are available in a Simple Aperture Photometry (SAP) or Pre-search Data Conditioning Simple Aperture Photometry (PDCSAP) formats. Because PDCSAP light curves are optimized to detect transit and eclipse signals, and remove other low-level variability that may be astrophysical in origin, we opted instead to use the SAP light curve. We normalized the SAP light curve by dividing by the median value long after the flare decay (MBJD~58605.6--58605.7). 

    \subsection{Zadko Telescope}
        The Zadko Telescope \citep{2017PASA...34....5C} is a 1\,m f/4 Cassegrain telescope situated in Western Australia. 
        For this campaign, its main instrument, an Andor IkonL camera, was replaced by a MicroLine ML50100 camera from Finger Lakes Instrumentation. The new camera allowed for a faster readout and thus a better temporal resolution of the light curve. The CCD is 8k$\times$6k, but we used it with a binning of 2$\times$2 to increase the signal-to-noise ratio, giving a pixel scale of $1.39\,\arcsec\,\text{pixel}^{-1}$.

        We used the Sloan $g^{\prime}$ filter (spanning $3885$--$5640$\AA) for the observations, which started as soon as possible after dusk, and lasted until Proxima Cen was too low on the horizon to safely operate the telescope. Each image was taken with a $10
        ,$s exposure, and needed 13.32\,s for the readout and preparation of the next exposure. Thus, our observation cadence for that night was about 23.3\,s. We took 20 bias and dark frames prior to the beginning of observations for calibration. During the observation night the conditions were good, with no visible
        clouds on the images. The stable temperature and calm wind allowed for fair observations.
        
        We applied standard photometric data reduction procedures using \texttt{ccdproc} \citep{2015ascl.soft10007C}. Twenty 10
        ,s dark exposures were bias corrected and median combined to remove effects of cosmic rays in the calibration frame. Bias and dark correction were applied to 180\,s flat-field exposures taken several nights after the observing campaign, and these flat-field frames were median combined. Raw target exposures were dark and bias corrected, and the corrected flat-field image was used to correct for spatial variations in the CCD sensitivity. Images were aligned using \texttt{astroalign} \citep{2020A&C....3200384B} to correct for slight drifts in the telescope pointing throughout the night.
        We used SExtractor \citep{1996A&AS..117..393B} to perform aperture photometry on Proxima Cen and nearby reference stars using a 12 pixel diameter ($16.7\,\arcsec$) aperture. We corrected the raw Proxima Cen light curve for systematic variations in flux over the course of the night by median normalizing the raw light curves of nearby neighboring stars around Proxima Cen, with similar colors ({Gaia} $B_p - R_p > 2$; $B_p - R_p = 3.796$ for Proxima Cen). We median combined these normalized light curves to produce a systematic reference light curve, and divided the Proxima Cen light curve by the systematic trend to produce a corrected light curve. Finally, we divided the Proxima Cen light curve by its median post-flare (MBJD~58605.6--58605.7) value to produce a light curve in relative intensity.

    
    \subsection{ANU 2.3m Telescope}

        We performed time-resolved spectroscopy with the Wide-Field Spectrograph (WiFeS) on board the ANU 2.3m Telescope at Siding Spring Observatory \citep{2007Ap&SS.310..255D}. WiFeS is an integral-field spectrograph that uses an image slicer with $25$ $1\,\arcsec\times 38\,\arcsec$ slitlets, resulting in a field of view of $25\,\arcsec \times 38\,\arcsec$. Conditions at the observatory on 2019 May 02 were poor, resulting in high levels of cloud absorption and intermittent observational coverage during rainy periods. Seeing varied between $1.3\,\arcsec$ and $2.0\,\arcsec$. We took spectra with the U7000 and R7000 gratings in the blue and red arms of the instrument, respectively, using the RT480 dichroic. We took 90\,s exposures of Proxima Cen using half-frame exposures, which resulted in a faster readout time of $\sim$30 s, and resulted in a reduced field of view of $12.5\,\arcsec \times 38\,\arcsec$. The corresponding wavelength ranges after calibration were $3500\,$\AA\--$4355\,$\AA\ and $5400\,$\AA\--$7000\,$\AA, at a resolution of $R \sim 7000$.
        
        For the calibration, we took bias, internal flat field (with a Quartz-Iodine lamp), wire, and arc frames (with a neon-argon lamp), using $5\times5\,\text{s}$ exposures for the red arm and $5\times90\,\text{s}$ exposures for the blue arm where appropriate. We took $3\times90\,\text{s}$ exposures of the standard star LTT4364 to calibrate for the bandpass response across wavelength for each grating.
        
        We used the PyWiFeS pipeline \citep{2014Ap&SS.349..617C} to derive and apply calibrations to each exposure. 
        To summarize the PyWiFeS calibration process, we performed overscan subtraction, bad pixel repair and cosmic ray rejection, and co-addition of bias frames, which we then subtracted from science frames. We co-added the internal flat-field frames, and derived wavelength and spatial ($y$-axis) zero-point solutions using the arc and wire frames respectively. 
        To derive the spectral flat-field response, we used the wavelength solution and the co-added flat-field frames, and applied the derived flat-field correction to each sky exposure. To derive the bandpass sensitivity response and flux scale, we extracted the spectrum of the standard star LTT4364 from a rectified ($x$, $y$, $\lambda$) grid, and determined the corrections by comparing the extracted spectrum to a spectro-photometrically calibrated reference spectrum.
        We applied the flux and bandpass corrections to each of the corrected Proxima Cen data cubes, before converting the telescope-frame ($x$, $y$, $\lambda$) cubes to the sky frame ($\alpha$, $\delta$, $\lambda$), where $\alpha$, $\delta$ denote J2000 R.A. and decl., respectively. To preserve the temporal resolution of our observation, we did not co-add any of the Proxima Cen exposures during the data reduction process. We extracted sky-subtracted spectra from the final calibrated spectral cubes using a custom script adapted from the internal PyWiFeS code.
        
        Due to cloudy conditions, accurately determining the continuum and line fluxes was not possible. We divided each exposure with the R7000 grating by the median value between $6000\,$ and $6030\,$\AA, selected because it was free of any prominent emission or absorption features that may vary significantly with flaring activity. Prominent [\ion{O}{1}] sky lines could not be readily subtracted due to the cloudy conditions. {Similarly, we divided each exposure with the U7000 grating by the median value between $4150$ and $4300$\,\AA. However, significantly higher cloud absorption in the $3500$--$4355$\,\AA U7000 band, and the intrinsically lower flux of Proxima Cen in this band means that the continuum estimates are even more uncertain than the R7000 band.}
        
        We used the \texttt{specutils} package\footnote{\url{https://specutils.readthedocs.io/}} to measure line equivalent widths in the extracted calibrated spectra.

    \subsection{Australian Square Kilometre Array Pathfinder (ASKAP)}
        We observed Proxima Cen with ASKAP \citep{2016PASA...33...42M} on 2019 May 02 09:00 UTC (scheduling block 8612) for 14 hours with 34 antennas, a central frequency of 888\,MHz with 288\,MHz bandwidth, 1\,MHz channels, and 10\,s integrations. We observed the primary calibrator PKS~B1934$-$638 for 30 minutes (scheduling block 8614) immediately following the Proxima Cen observation. 
        {To calibrate the frequency-dependent $XY$-phase, an external noise source (the ``on-dish calibrator'' system) is employed on each of the ASKAP antennas. This system measures the $XY$ phase of each dual-polarization pair of phased-array feed beams, and adjusts the phase of the $Y$-polarization beamformer weights so that the $XY$ phase approaches zero. Further details of the ASKAP on-dish calibration system are provided in \citet{chippendale2019} and \citet{hotan_askap}.}
        
        We reduced the data using the Common Astronomy Software Applications (\textsc{casa}) package version~{5.3.0-143} \citep{casa}. We used PKS~B1934$-$638 to calibrate the flux scale, the instrumental bandpass, and polarization leakage.  We performed basic flagging to remove radio-frequency interference that affected approximately 20\% of the data, primarily from known mobile phone bands. 
        
\begin{figure}
        \centering
        \includegraphics[width=1.0\linewidth]{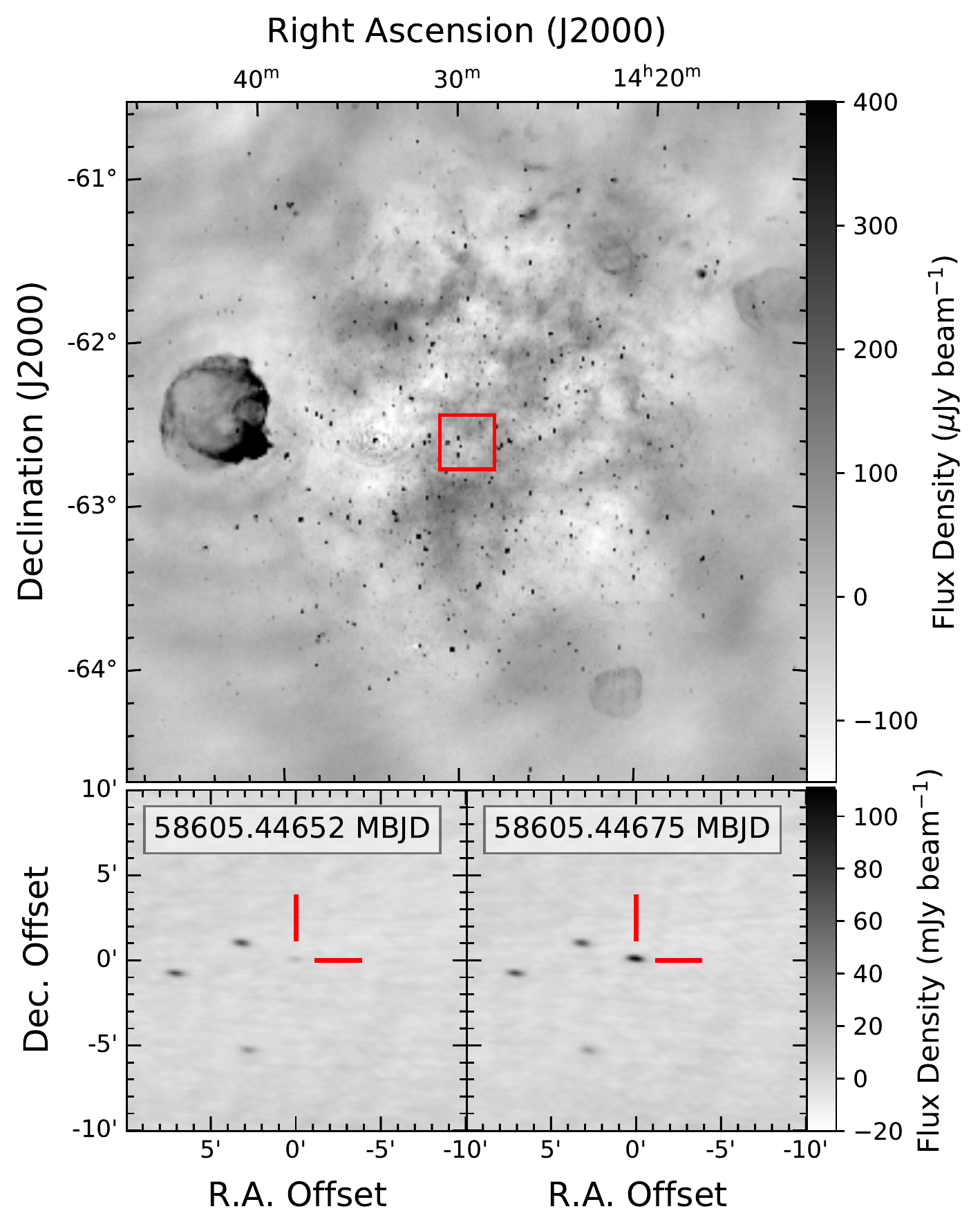}
        \caption{{ASKAP Stokes $I$ continuum images. Top panel: overview of the Proxima Cen field, imaged over the 14-hour observation. The grayscale intensity ranges from $-150$ to $400\,\mu\text{Jy}\,\text{beam}^{-1}$. This shows the presence of multiple strong point and extended sources, along with diffuse galactic emission. Bottom panel: 10\,s deconvolved snapshot images of the local $20\arcmin \times 20\arcmin$ region around Proxima Cen, taken 20\,s before (58605.44652 MBJD; left), and around the peak of AB1 (58605.44675 MBJD; right). The region shown in the snapshot images is indicated by the red square in the top sub-figure. The grayscale intensity ranges from $-20$ to $110$\,mJy\,beam$^{-1}$ for the 10\,s snapshot images.}}
        \label{fig:askap_image}
        \end{figure}

        We used a mask excluding a $4\arcmin$ square region centered on Proxima Cen to allow modeling of the field sources without removing the time- and frequency-dependent effects of Proxima Cen. {The large exclusion window around the target also ensured that the flux density present in strong point-spread function (PSF) sidelobes during burst events were not modeled as point sources and removed during deconvolution, ensuring that its total flux density was preserved.} We used the task \textsc{tclean} to perform deconvolution with a Briggs weighting and a robustness of 0.0. We used the \textsc{mtmfs} algorithm (with scales of 0, 5, 15, 50, and 150 pixels and a cell size of $2.5\,\arcsec$) to account for complex field sources, and used two Taylor terms to model sources with non-flat spectra. We imaged a $6000\times6000$ pixel field ($250\arcmin \times 250\arcmin$) to include the full primary beam and first null, and deconvolved to a residual of $\sim$3\,mJy\,beam$^{-1}$ to minimize PSF side-lobe confusion at the location of Proxima Cen. We subtracted the field model from the visibilities with the task \textsc{uvsub}, and vector-averaged all baselines greater than 200\,m to generate the dynamic spectra for each of the instrumental polarizations. {We show a deconvolved image of the full $250\arcmin$ field over the 14
        ,hr ASKAP observation in Figure \ref{fig:askap_image}.} 
        
        We formed dynamic spectra for the four Stokes parameters ($I$, $Q$, $U$, $V$) following \citet{2019MNRAS.488..559Z}, such that they were consistent with the IAU convention of polarization. To improve the signal-to-noise ratio in the dynamic spectra, we averaged the Stokes $I$ and $V$ products by a factor of 3 in frequency to produce final dynamic spectra with a resolution of 3\,MHz in frequency and 10\,s in time. Similarly, we averaged the Stokes $Q$ and $U$ dynamic spectra by a factor of 6 in frequency, giving these products a resolution of 6\,MHz in frequency and 10\,s in time. We produced radio light curves by averaging the dynamic spectra across frequency. The rms sensitivity of our dynamic spectra is 12\,mJy for Stokes $I$, and $11$\,mJy for Stokes $Q$, $U$, and $V$, calculated by taking the standard deviation of the imaginary component of the dynamic spectra visibilities. In a similar way, we calculated the rms sensitivity of the light curves, finding 1.4\,mJy for Stokes $I$, and 1.1\,mJy for Stokes $Q$, $U$, and $V$.

\section{Detection of Flare and Radio Burst Events}
\label{multiwave_results}

\begin{figure*}
    \centering
    \includegraphics[width=\linewidth]{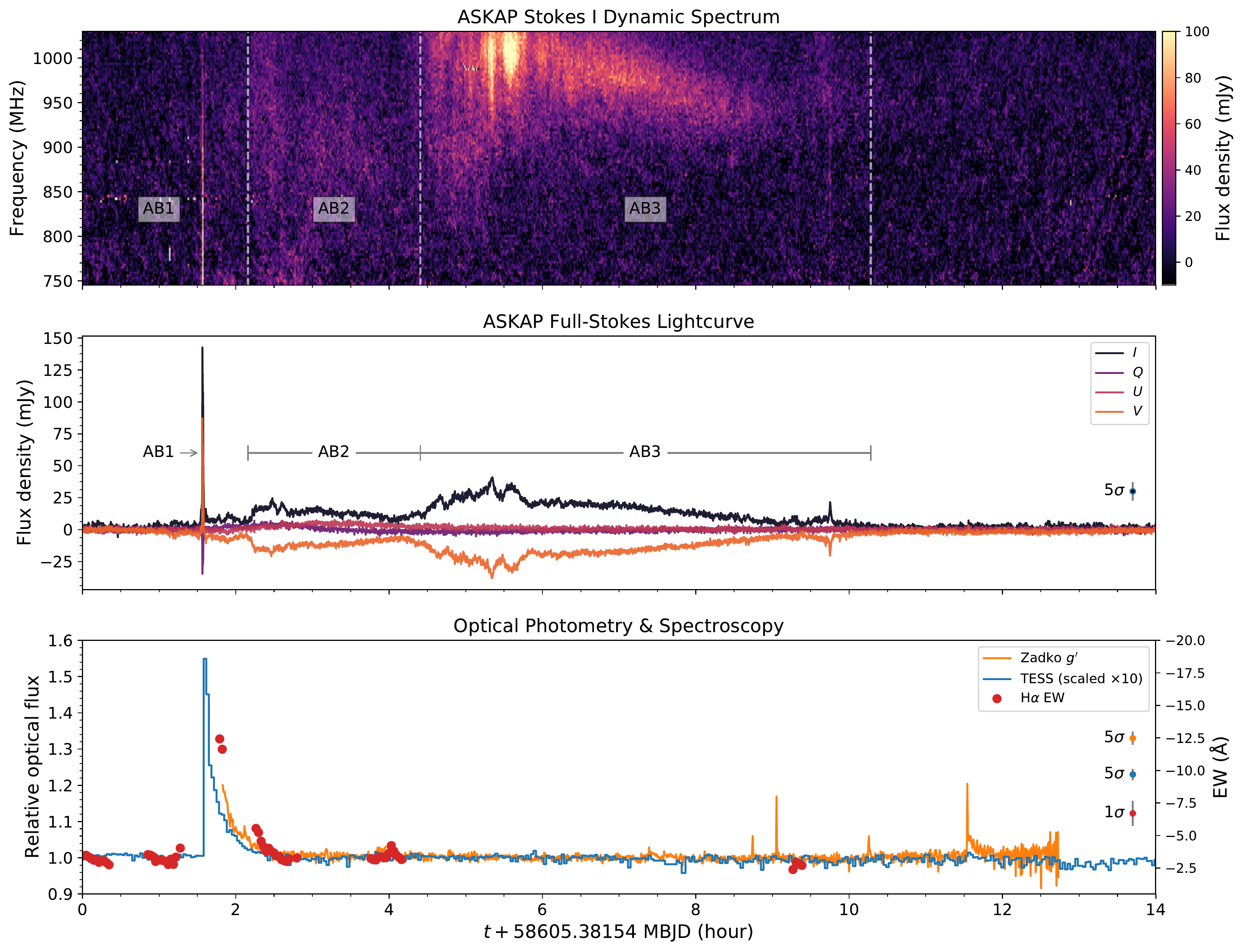}
    \caption{Multiwavelength overview of 2019 May 2 observations. {Top panel}: ASKAP Stokes $I$ dynamic spectrum, showing intensity as a function of frequency and time. The resolution is 10\,s in time and 3\,MHz in frequency. The short timescale and broadband morphology of AB1, complex morphology of AB2, and long-duration, slow-drift morphology of AB3 are evident. Dashed vertical lines indicate the time intervals for AB2 and AB3. The burst labels AB1, AB2, and AB3 are indicated in the figure. {Middle panel} ASKAP light curves in Stokes $I$, $Q$, $U$, and $V$, colored in purple-black, dark purple, light purple, and orange, respectively. Burst labels are as in the top panel. {Bottom panel}: median-normalized photometric light curves from TESS (blue curve) and the Zadko Telescope $g^\prime$ band (orange curve), both with values shown on the left abscissa, and H$\alpha$ equivalent width from WiFeS on board the ANU 2.3m Telescope (red dots, values shown on right abscissa). Gaps in the equivalent width measurements are due to poor weather at the Siding Spring Observatory. For visual clarity, the TESS light curve has been scaled by a factor of 10. Typical uncertainties are indicated in the figure on the right.}
    \label{fig:multiwavelength}
\end{figure*}

    \subsection{Photometric Flare Detection, Energy, and Temporal Modeling}
    We present the photometric light curves in the bottom panel of Figure \ref{fig:multiwavelength}. These observations show the large, long-duration flare on MBJD~58605, which had a duration of approximately 1\,hr. 
    
    
    We determined the radiated flare energy as follows. The flare energy $E_{f,p}$ through a photometric passband $p$ is given by $E_{f,p} = \text{ED}\times L_{p}$, where $\text{ED}$ is the `equivalent duration' of the flare, and $L_{p}$ is the total quiescent luminosity of the star in passband $p$. The equivalent duration is given by
    \begin{equation}
        \text{ED} = \int_{t_0}^{t_1} \left( \frac{I_f(t) - I_0}{I_0} \right)dt,
    \end{equation}
    which we evaluate using Simpson's rule. Here, $t_0$ and $t_1$ are the start and end times of the flare, estimated by visual inspection of the light curve, $I_f(t)$ is the intensity of the flare as a function of time, and $I_0$ is the median quiescent intensity. To compute the quiescent luminosity of the star, we obtained the flux-calibrated spectrum of Proxima Cen presented in \citet{2017A&A...603A..58R}, multiplying by $4\pi d^2$ to obtain the spectral luminosity $L_{\lambda}$. We estimate the total quiescent luminosity through passband $p$ as $L_p \approx \langle {L}_{\lambda,p}\rangle \Delta \lambda_p$, where $\langle {L}_{\lambda,p}\rangle$ is the mean quiescent spectral luminosity in passband $p$, and $\Delta \lambda_p$ is the bandwidth of passband $p$. We evaluated the mean quiescent spectral luminosity using
    \begin{equation}
        \langle{L}_{\lambda,p} \rangle = \frac{\int_0^\infty L_{\lambda}T_p(\lambda)\lambda d\lambda}{\int_0^\infty T_p(\lambda)\lambda d\lambda},
    \end{equation}
    where $T_p(\lambda)$ is the filter transmission curve for passband $p$.
    This quantity is independent of any global scaling factors of the transmission curve, enabling a more reliable estimate of the total quiescent luminosity through passband $p$. To measure the the bandwidth $\Delta \lambda_p$, we computed the equivalent rectangular width,
    \begin{equation}
        W = \frac{\int_0^\infty T_p(\lambda) d\lambda}{\text{max}(T_p(\lambda))},
    \end{equation}
    which is the width of a rectangle of height 1 and area equal to the total area beneath the filter transmission curve $T_p(\lambda)$.
    Applying these calculations to the TESS and Sloan $g^{\prime}$ passbands, we obtain quiescent luminosities of $1.1\times10^{30}\,\text{erg}\,\text{s}^{-1}$ and $2.3 \times 10^{28}\,\text{erg}\,\text{s}^{-1}$ respectively. TESS observations cover the full duration of the flare, albeit with relatively low temporal resolution (2\,minute cadence), enabling us to compute a TESS-band flare equivalent duration of $31.2\pm 0.3$\,s, yielding a TESS-band flare energy $E_{f,\mathrm{TESS}} = 3.38 \pm 0 .03 \times 10^{31}\,\text{erg}$. 
    To estimate the bolometric flare energy, we follow \citet{2015ApJ...809...79O} and adopt a 9000\,K flare blackbody profile, and evaluate the fraction of energy radiated within the TESS and Sloan $g^\prime$ passbands, respectively, finding 
    $E_{f,\mathrm{TESS}}/E_{\mathrm{bol}} = 0.21$ and
    $E_{f,g^\prime}/E_{\mathrm{bol}} = 0.22$. 
    Using the TESS observation, which covers the full duration of the flare, we calculate a bolometric flare energy of $1.64 \pm 0.01 \times 10^{32}\,\text{erg}$.
    {We note that the quoted uncertainties in this section are statistical only, and factors such as the light-curve normalization, accurate determination of quiescent luminosity, and choice of flare blackbody temperature introduce systematic uncertainties, which we estimate contribute to errors on the order of 10\% of the quoted values.}

\begin{figure*}
    \centering
    \includegraphics[width=\linewidth]{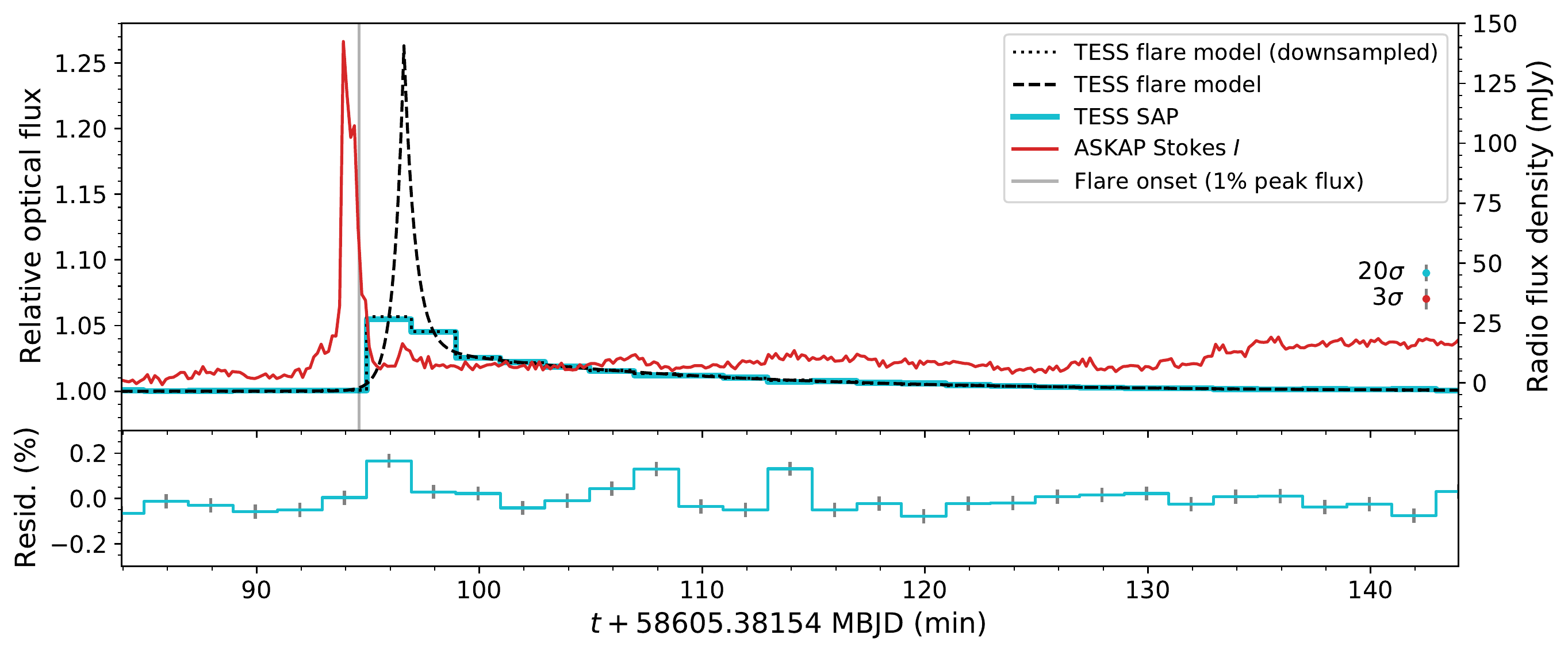}
    \caption{{Top}: relative timing of ASKAP Burst 1 (AB1), shown in red, and the main optical flare observed with TESS, shown in blue. The black dashed curve shows the best-fitting empirical model to the flare observed by TESS. 
    The black dotted line shows the best-fitting model down-sampled onto the 2\,minute resolution time series of the TESS observations. The  vertical gray line shows the flare onset time, at which the intensity first reaches $1\%$ of the flare peak. The right abscissa shows the ASKAP flux densities, and the left abscissa shows the relative flux from TESS-band photometry. Typical uncertainties are shown toward the figure right, upscaled for visual clarity. The radio peak and the estimated flare onset are separated by 42\,s. {Bottom}: percentage residuals after subtracting the best-fitting flare model from the TESS light curve. Error bars on the residuals are $1\sigma$.}
    \label{fig:flare_timing}
\end{figure*}

    The 2\,minute cadence of the TESS light curve undersamples the impulsive rise phase of the flare. To obtain a more informed estimate of the flare onset time, we modeled the temporal morphology of the flare using the following piecewise model, adapted from the empirical flare template presented in \citet{2014ApJ...797..122D}:
    \begin{align}
        \Delta I &= \frac{I_f - I_0}{I_0}\\
            &= A\times \left\{
            \begin{array}{lc}
                 A_0^{({t - t_p})/({t_0 - t_p})} & t < t_p \\
                (1.0 - A_g) e^{({t-t_p})/{\tau_i}} \\ + A_g e^{ ({t-t_p})/{\tau_g}} & t \geq t_p\\
            \end{array} \right. \text{,}
    \end{align}
    where $A$ is the peak fractional amplitude above quiescence, $t_p$ is a the time at flare peak relative to the ASKAP observation start time MBJD~58605.38154, $A_0$ is the relative pre-flare amplitude in the last TESS exposure at a set time MBJD~$t_0 = 58605.44678$ (93.94 minutes after the beginning of the ASKAP observation at MBJD 58604.38154),
    $\tau_i$ is the impulsive decay timescale, and $A_g$ and $\tau_g$ are the gradual decay amplitude and timescales. {We model the impulsive rise with an exponential rather than the quartic model of \citet{2014ApJ...797..122D} to reduce the number of free parameters to determine from the low-resolution TESS light curve.}
    We used \texttt{emcee} to estimate the parameters and sample their posterior probability distributions, first evaluating each candidate model on a high-resolution time grid before resampling the candidate model onto the lower-resolution (2\,minute cadence) TESS time series. We set uniform priors on each parameter, constraining $A_0 > 0$, $\tau_i, \tau_g < 0$, and $A>0.03$. The resulting central parameter estimates and 64\% confidence limits are as follows: $t_p = 96.61^{+0.23}_{-0.17}$\,minutes, $A = 0.26^{+0.19}_{-0.08}$, $A_0 = 2.2^{+1.9}_{-1.3}\times 10^{-3}$, $A_g = 0.13^{+0.06}_{-0.05}$, $\tau_g = -12.6\pm 0.2$\,minutes, $\tau_i = -0.5\pm{0.2}$\,minutes. The best-fitting model is shown in Figure \ref{fig:flare_timing}, alongside the photometric light curve from TESS and the radio light curve from ASKAP.
    
    \subsection{WiFeS Time-Resolved Spectroscopy}

\begin{figure*}
    \centering
    \includegraphics[width=\linewidth]{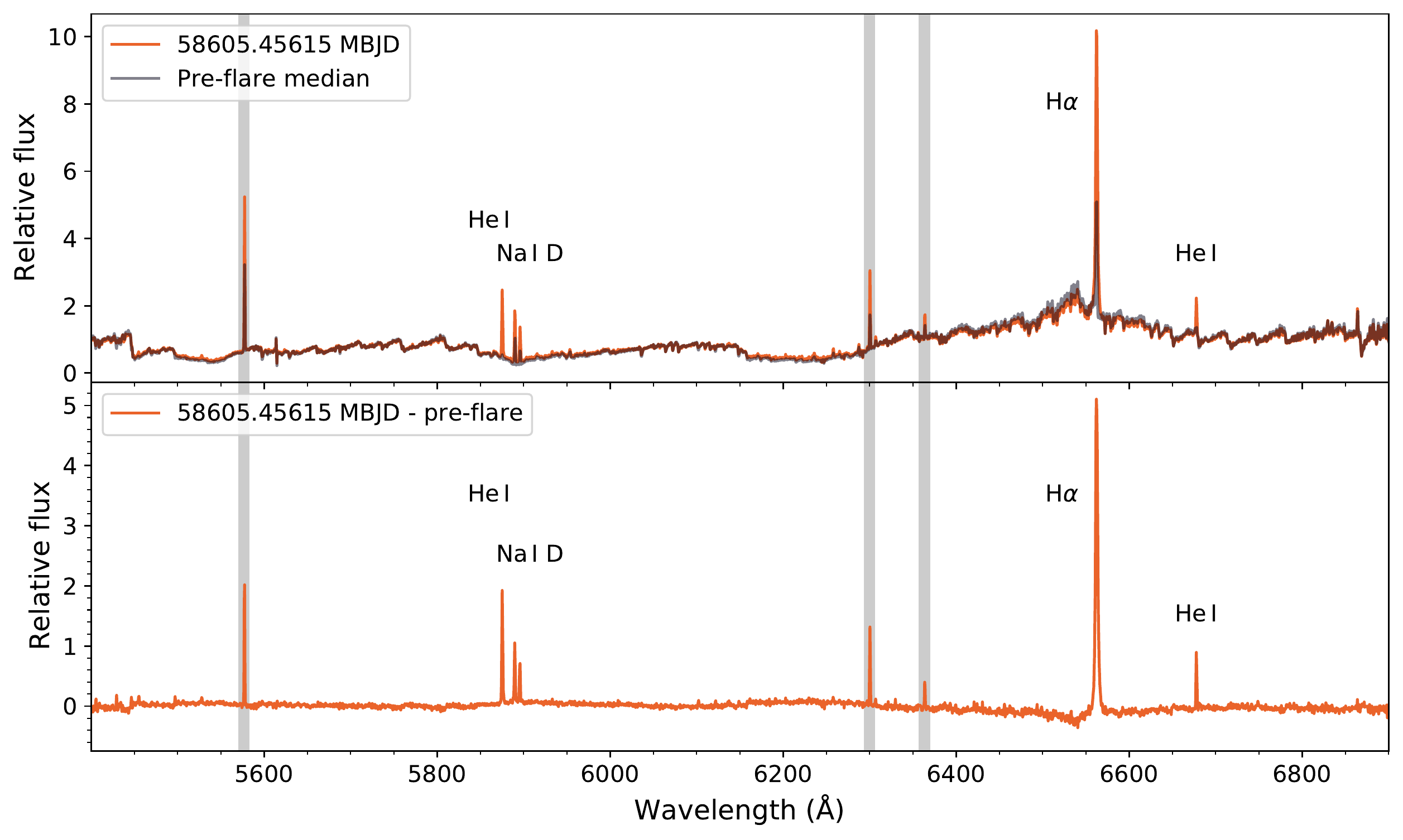}
    \caption{\textit{Top}: WiFeS R7000 ($\lambda5400\text{--}7000$\,\AA) normalized spectra, showing a median of 10 exposures before the main flare (gray curve), and one 90\,s exposure taken at MBJD~58605.45615 during the decay phase of the flare (orange curve). \textit{Bottom}: quiescent-subtracted flare decay spectrum. Brightening of chromospheric emission lines, such as H$\alpha$, He\,I, and the Na\,I doublet (labelled in diagram) are clearly visible. Due to cloud cover, atmospheric [O~I] lines at $5577$, $6300$, and $6363$\,\AA~, indicated in gray-shaded regions, could not be subtracted and should be ignored.}
    \label{fig:flare_spec}
\end{figure*}

\begin{figure*}[ht!]
        \centering
        \includegraphics[width=\linewidth]{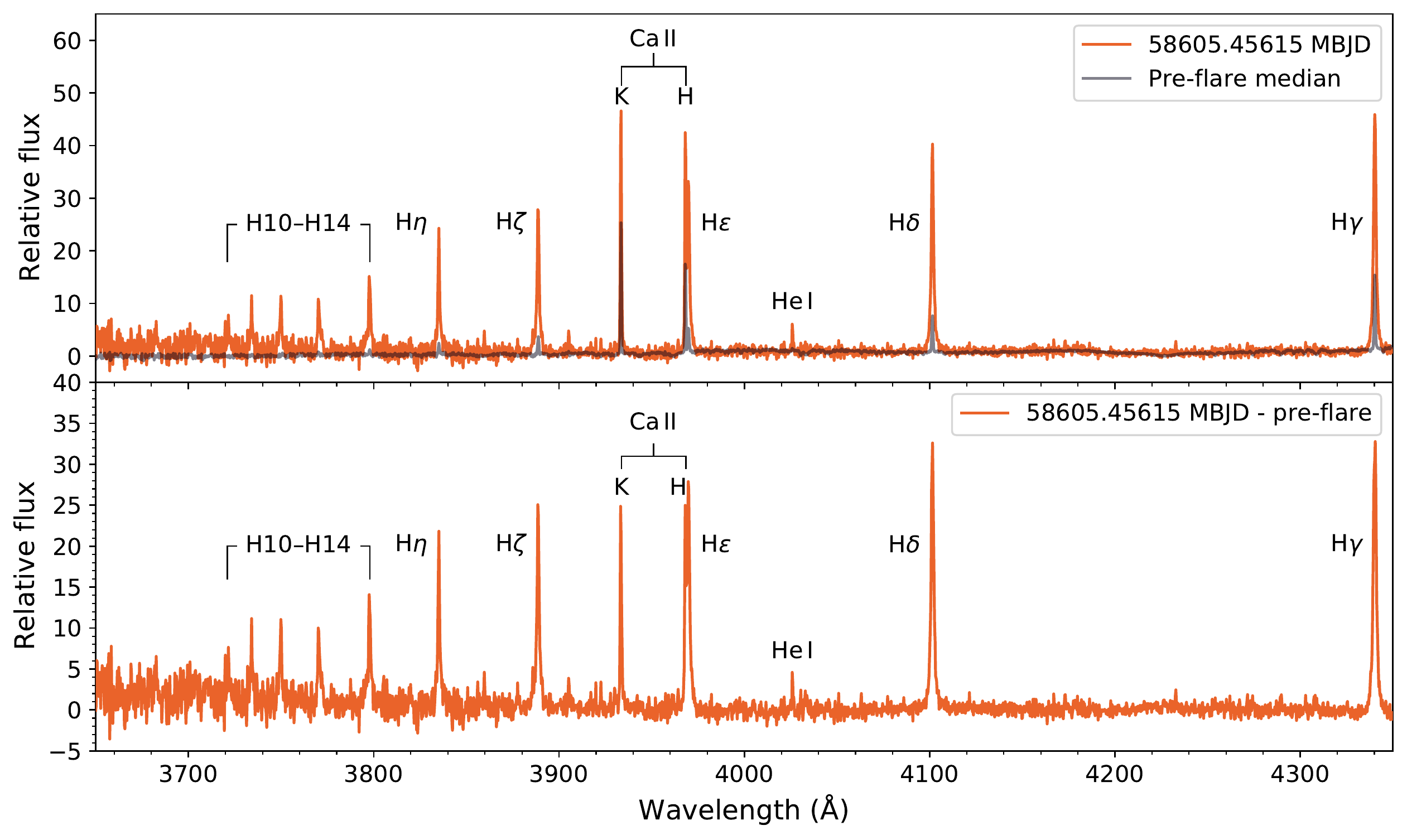}
        \caption{Same as Figure \ref{fig:flare_spec}, but for the WiFeS B7000 ($\lambda3500\text{--}4355$\,\AA) spectra. Brightening and broadening of Balmer lines, from H$\gamma$ through to H14 are evident, along with an enhancement of the continuum toward the blue end of the spectrum.}
        \label{fig:flare_bspec}
    \end{figure*}

    Spectroscopic monitoring with the WiFeS on board the ANU 2.3m Telescope \citep{2007Ap&SS.310..255D} 
    shows broadening and intensification of the Balmer lines, and the appearance of other chromospheric emission lines (e.g., He~I $\lambda\lambda 4026 ,5876, 6678$\,\AA) during this flare (Figures \ref{fig:flare_spec} and \ref{fig:flare_bspec}). This indicates substantial energy deposition by accelerated particles into the stellar chromosphere. Unfortunately, these observations were affected by poor weather,
    making it difficult to reliably detect variability in chromospheric emission line strength over the whole night. Along with the appearance of higher-order Balmer lines (up to H14 $\lambda 3721$\,\AA), Figure \ref{fig:flare_bspec} also shows a continuum enhancement toward the blue end of the spectrum. Figure \ref{fig:multiwavelength} shows that the equivalent width of the H$\alpha$ line is well correlated with the gradual decay of the flare continuum, consistent with previous observational results \citep[e.g.][]{1991ApJ...378..725H}.
    
    \begin{deluxetable*}{ccccccccc}[ht]
        \centerwidetable
        \tablecaption{Properties of the radio bursts.\label{tab:radio_burst}}
        \tablehead{
        \colhead{Label }  & \colhead{$t + \text{MBJD}~58605$} & \colhead{$\Delta t$} & \colhead{$S_\text{peak}$}  & \colhead{$T_{b,\text{peak}}$\,\tablenotemark{\footnotesize{a}}}  & \colhead{$f_L$} & \colhead{$f_C$}  & \colhead{$f_P$}  & \colhead{$\dot{\nu}$}\\
        \colhead{}        & \colhead{(days)}                      & \colhead{(minutes)}      & \colhead{(mJy)}             & \colhead{($\times 10^{11}$\,K)}     & \colhead{(\%)}     & \colhead{(\%)}    & \colhead{(\%)}   & \colhead{(MHz\,s$^{-1}$)}}
        \startdata
             AB1 & $0.4466$      & $1.3$ & $142.4\pm 1.4$ & $2.93 \pm 0.03$ & $36.5 \pm 1.1$ & $-52.1 \pm 1.3$ & $57.1 \pm 1.4$ & $-4.70 \pm 0.66$ \\
             AB2 & $0.4715$      & $119$ & $24.7 \pm 1.4$ & $0.51 \pm 0.02$ & $43.3 \pm 0.4$  & $87.4 \pm 0.8$  & $98.5\pm 1.0$ & Complex \tablenotemark{\footnotesize{b}} \\
             AB3 & $0.5652$      & $354$ & $41.0 \pm 1.4$ & $0.84 \pm 0.02$ & $<9$\,\tablenotemark{\footnotesize{c}} & $91 \pm 4$ & $91 \pm 4$ & $5.9\pm 0.3\times 10^{-3}$ \\ 
        \enddata
        \tablenotetext{\footnotesize{a}}{This is a lower limit owing to the conservatively large size of the emission region used.}
        \tablenotetext{\footnotesize{b}}{AB2 exhibits reversals in its drift direction, preventing simple characterization with a linear drift rate.}
        \tablenotetext{\footnotesize{c}}{Some linear polarization is evident at the beginning of AB3. However, this is likely to be from the overlapping tail end of AB2.}
        \tablecomments{Columns, from left to right: burst label, start time, duration, peak flux density, brightness temperature; fractional linear polarization; circular polarization, and total polarization, and frequency drift rate. {To avoid positive biases due to Ricean statistics, fractional polarizations were measured by masking out insignificant values (signal-to-noise ratio $<1$) in the dynamic spectra, before averaging over frequency and taking the quadrature sum where relevant.}}
    \end{deluxetable*}

    \subsection{Radio Bursts Detected with ASKAP}


    \begin{figure}
        \centering
        \includegraphics[width=\linewidth]{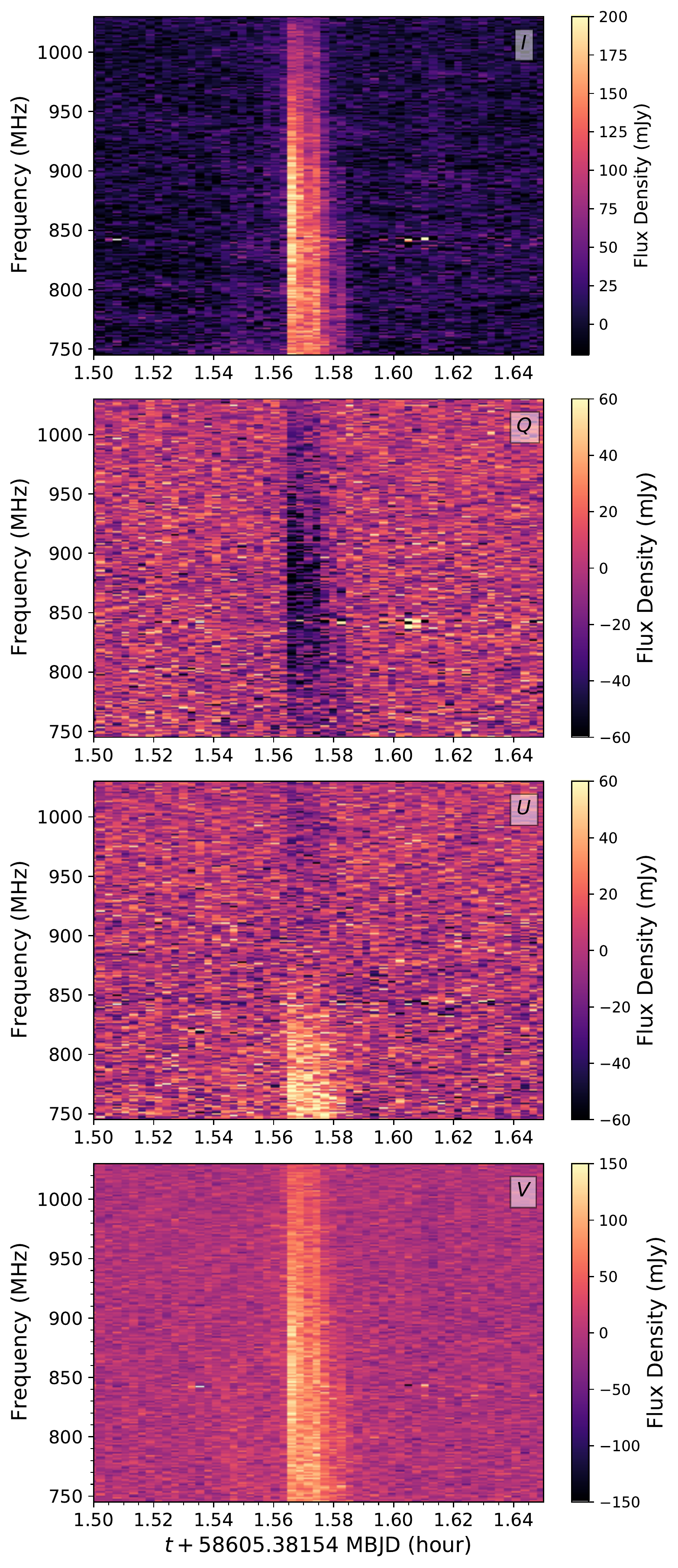}
        \caption{Dynamic spectrum of AB1 in Stokes $I$, $Q$, $U$ and $V$ (top to bottom), showing the rapid onset (rise time $\lesssim 10\,\text{s}$) linear polarization, and downward-drifting upper envelope of AB1. The temporal and spectral resolution are 10\,s and 1\,MHz, respectively.}
        \label{fig:askap_ab1_ds}
    \end{figure}
    
    \begin{figure*}
        \centering
        \includegraphics[width=\linewidth]{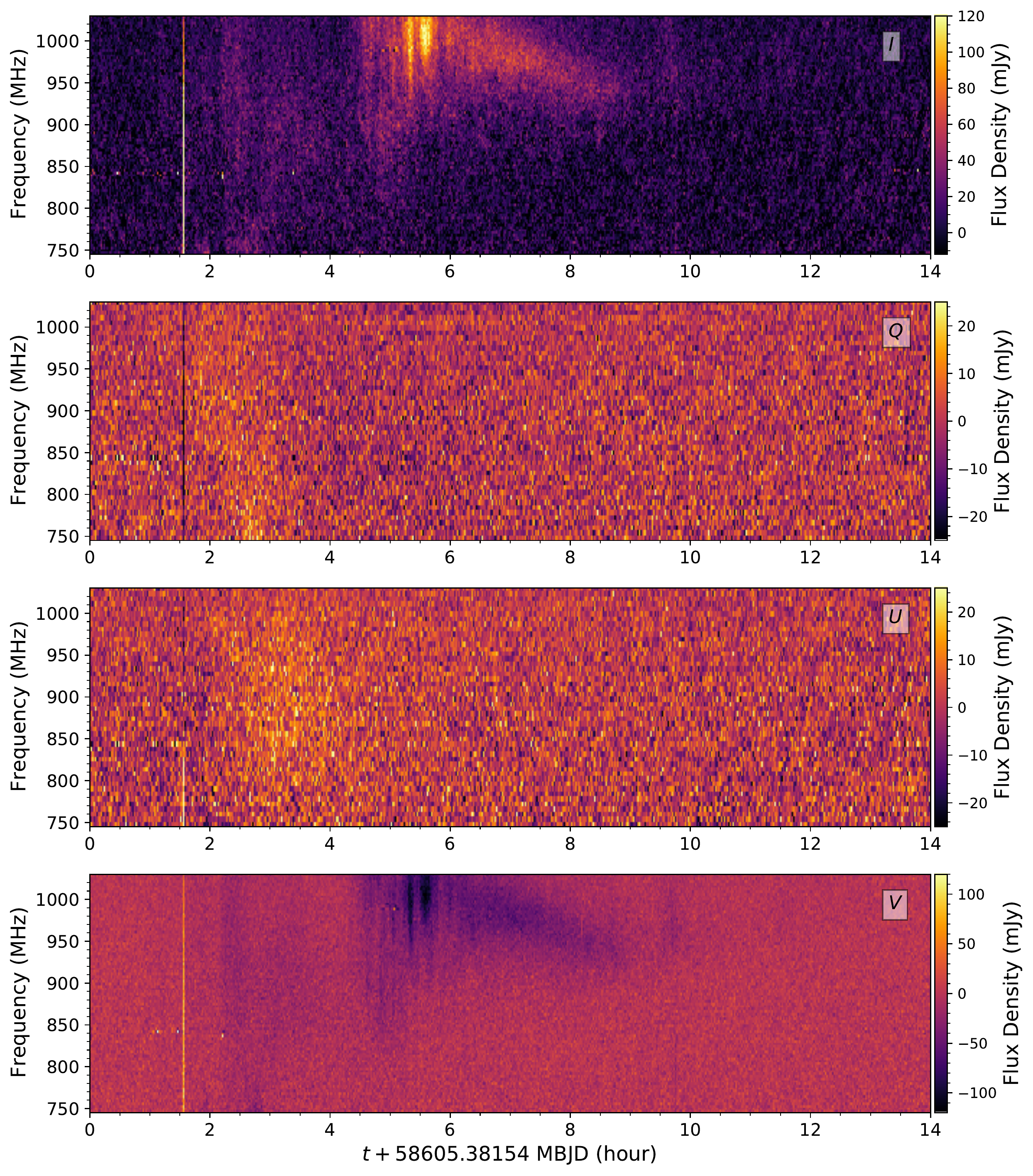}
        \caption{Full-Stokes ASKAP dynamic spectrum, showing intensity as a function of frequency and time for Stokes $I$, $Q$, $U$, and $V$ from top to bottom. The high degree of circular polarization for all bursts, and diffuse spectro-temporal structure of the linearly polarized component of AB2 are evident. The spectral resolution is 3\,MHz for Stokes $I$ and $V$, and 6\,MHz for Stokes $Q$ and $U$. The temporal resolution is 10\,s for all Stokes parameters.}
        \label{fig:askap_stokes_ds}
    \end{figure*}

    Observations with the ASKAP \citep{2016PASA...33...42M}) 
    centered at 888\,MHz show three successive bursts, which we denote as ASKAP Bursts 1, 2, and 3 (AB1, AB2, and AB3 respectively). Figure \ref{fig:multiwavelength} shows the total intensity (Stokes $I$) dynamic spectrum and full-Stokes light curves in their multiwavelength context, and the full-Stokes dynamic spectra of the bursts are shown in Figures \ref{fig:askap_ab1_ds} and \ref{fig:askap_stokes_ds}. {Figure \ref{fig:askap_image} shows two 10\,s snapshot images, taken 20\,s prior to (bottom left) and at the peak of AB1 (bottom right)}. 
    Table \ref{tab:radio_burst} summarizes the important burst properties. The important observational characteristics of these bursts include high flux densities (peak flux density $> 100\,\text{mJy}$, high degree of circular polarization ($|V/I| > 50\%$), presence of moderate degrees of linear polarization for AB1 and AB2 ($\sim 40\%$), and narrow-band and sharp-cutoff spectral features. {To measure fractional polarizations, we masked insignificant values (signal-to-noise ratio $< 1$) in the dynamic spectra for each Stokes parameter, before averaging over frequency and taking quadrature sums where relevant. This mitigated positive biases toward the fractional polarization due to the Ricean statistics of quadrature-summed signals.}
    
    To measure the drift rate of ASKAP Bursts 1 and 3 (AB1 and AB3), we measured a characteristic frequency of the burst for each temporal integration. We fit the time-frequency ordinate pairs with a linear model, using \texttt{emcee} \citep{2013PASP..125..306F} to estimate the slope and intercept parameters. AB1 is broadband, and its lower spectral cutoff is below the frequency range of the observations. For this reason, we measured the upper frequency cutoff (defined as the highest frequency where the intensity exceeds a $6\sigma$ threshold of $71$\,mJy), and took this as the characteristic frequency. The slope in the upper frequency cutoff of AB1 is evident in Figure \ref{fig:askap_ab1_ds}. For AB3, we took the frequency of maximum flux density for each integration as the characteristic burst frequency, since the drifting component of AB3 is contained within the bandwidth of our observation. The derived drift rates are $-4.7\pm 0.7\,\text{MHz}\,\text{s}^{-1}$ for AB1 and $-7.0\pm0.2\,\text{kHz}\,\text{s}^{-1}$ for AB3.
    
    
    
    \subsection{Determining the Radio Emission Mechanism}
    \label{subsec:emission_mech}
    The brightness temperature is an important diagnostic of the emission mechanism, with very high brightness temperatures indicating a coherent emission process. 
    Conservatively assuming a source size equal to the size of the full stellar disk {($R_* = 0.146 R_\odot$; \citealp{2017A&A...603A..58R}),} we can calculate a lower limit on the brightness temperature using Equation 14 from \citet{1985ARA&A..23..169D}. The lower limits on the brightness temperature for each burst are given in Table 1, and lie within the range of $10^{10}$--$10^{11}$\,K. Due to the high brightness temperatures, the high degrees of polarization (up to 100\%), and the spectral structure of the bursts (including narrow-band features and sharp frequency cutoffs), coherent emission is strongly favored. {In the context of stellar radio emission, the most plausible coherent emission mechanisms are the electron cyclotron maser instability (ECMI), or plasma emission \citep{2017RvMPP...1....5M, 1985ARA&A..23..169D}}. 
    
    In the case of AB3, assuming that the drifting component with bandwidth $\Delta \nu \approx 120\,\text{MHz}$ arises from a single ECMI source, we can estimate the length of the emitting region $L_s$ as $L_s \approx L_B \Delta \nu / \nu$, where $L_B = | B / \nabla B |$ is the magnetic scale height, taken to be close to the stellar radius (i.e. $\sim1\times 10^{10}$\,cm). For $\Delta \nu = 120\,\text{MHz}$, $\nu = 950\,\text{MHz}$, we obtain a source length of $1.3\times 10^{9}\,\text{cm}$. Using this as the characteristic source size, we obtain brightness temperatures in the range of $10^{13}$\,K to $10^{14}$\,K, confirming the need for a coherent emission mechanism.
    
    AB1 and AB2 also exhibit elliptical polarization, as can be seen in Figures \ref{fig:askap_ab1_ds} and \ref{fig:askap_stokes_ds}.
    Possible interpretations of its origin include intrinsically elliptically polarized emission arising from a strongly rarefied duct \citep{2019MNRAS.488..559Z, 1991A&A...249..250M}; or coupling of the extraordinary and ordinary magneto-ionic modes as the radiation traverses an inhomogeneous quasi-transverse region \citep{1963PASJ...15..195K}. In the former case, only the ECMI can be responsible for the emission, while in the latter case both fundamental plasma emission or ECMI emission are plausible emission mechanisms. While both mechanisms are plausible, reversals in the frequency drift of AB2 and the lower frequency cutoff of AB3 can be more readily explained by the modulation and/or geometrical beaming effects of the ECMI mechanism, whereas plasma emission would require upward and downward motion and/or density fluctuations of an excited plasma region.
    
    \subsection{Inferring Physical Properties from the ECMI Emission}
    
    The properties of AB2 and AB3, which make up the type IV burst, can be used to infer properties of the associated post-eruptive loop system.
    The $\sim 40\%$ degree of linear polarization in AB2 implies a strongly under-dense source region, {since Faraday rotation effects, such as differential Faraday rotation and bandwidth depolarization, should be strong in a dense and highly magnetized stellar corona \citep{1998MNRAS.299..189S,1991A&A...249..250M}. \citet{1991A&A...249..250M} considered the elliptical polarization of Jovian decametric radio emission \citep{1991A&A...251..339L,1991A&A...247..235B}, and argued that for the elliptical polarization to be preserved along the propagation path, mode coupling around the emission region, and along the propagation must be weak. This leads to a condition on the electron density,
    \begin{equation}
        n_e \lesssim \alpha (\nu / 25\,\text{MHz})\,\text{,}
    \end{equation}
    where $\alpha$ is a geometrical factor of order unity, which we take to be $1$. Using this, we compute an electron density upper limit of $\sim 30\,\text{cm}^{-3}$, corresponding to our lowest observing frequency $\nu = 744$\,MHz. This limit is derived based on the emission angle relative to the magnetic field $\theta$ satisfying $\cos^2(\theta) \ll 1$, along with assumptions that $\nu_p^2 / \nu^2 \ll 1$ and $\nu_p^2 \sin(\theta)/2 \nu^2 \ll 1 - \nu_c / \nu \ll 1$ close to the emission region, and $\nu_p^2/\nu^2 \ll 1$, $\nu_c / \nu \ll 1$ more than about one magnetic scale length from the emission region \citep{1991A&A...249..250M}.}
    The emission frequency for ECMI is close to the local electron cyclotron frequency $\nu_c = e B / m_e c \approx 2.8 B$\,MHz, or its second harmonic. Assuming first harmonic emission, ECMI radiation in the ASKAP band (744--1031\,MHz) corresponds to local magnetic field strengths ranging from $\sim$270--370\,G. The low plasma density inferred from the elliptical polarization implies a local plasma frequency $\nu_p \leq 100\,\text{kHz}$, a factor of at least $\sim 1000$ lower than the local electron cyclotron frequency. These properties satisfy the condition $\nu_p / \nu_c \ll 1$ for operation of the ECMI \citep{1982ApJ...259..844M}.
    
    Using the observed frequencies of the emission, we can estimate the vertical extent of the post-eruptive loop system. The average surface magnetic field of Proxima Cen is $600\pm150$\,G \citep{2008A&A...489L..45R}. Considering that the magnetic field strength in active regions may be substantially stronger, we use a range of basal magnetic field strengths from 600--1600\,G. Assuming (1) that the ECMI-emitting region fills a fraction between 0.1 and 0.5 of the length active loop (consistent with the fractional bandwidth of the bursts) from the footpoints, (2) an upper frequency cutoff for AB3 of $\sim 1200$\,MHz, and (3) a dipolar loop geometry (so that $B(r) \propto r^{-3}$), we estimate loop sizes from $\sim$0.3--1.0 stellar radii ($\sim 0.3$--$1.0\times 10^{10}\,$cm).

\subsection{Probability of a Coincident Flare and Radio Burst}

    To compute the probability that the optical flare and AB1 are temporally aligned by chance coincidence, we followed the approach of \citet{2005ApJ...621..398O}. We assume the probability distribution function of observing $N$ flares occurring at a rate $\lambda$ within a time interval $\Delta t$ is 
    \begin{equation}
    \label{eq:prob_coinc}
        P(N~|~\lambda, \Delta t) = e^{- \lambda \Delta t}( \lambda \Delta t)^{N}/N!\,\text{.}
    \end{equation}
    If two flare events in disparate wave bands are independent, then the probability of observing both within a time interval $\tau$ of each other is just the product of their independent rate probabilities. For the optical flare, using the bolometric flare energy of $\sim 1.6 \times 10^{32}$\,erg we calculate a rate for flare of this energy and higher of $\sim 0.04\,\text{day}^{-1}$ using the cumulative flare frequency distribution model from \citet{2018ApJ...860L..30H}. The burst rate of Proxima Cen at radio frequencies is not well constrained and is likely to be frequency dependent \citep{2019ApJ...871..214V}. Given the six bursts observed in 46\,hr of observing time with ASKAP, we estimate an ASKAP-band radio burst rate of $\sim 3\,\text{days}^{-1}$, above a detection threshold of a few millijansky. 
    
    Using the flare temporal model described above, we determine that the time at which the flare reaches 1\% of its peak intensity is $94.59^{+0.32}_{-0.27}$\,min. Taking this as the flare onset time, we obtain a temporal offset between the peak of AB1 and the flare onset of $42^{+19}_{-17}$\,s, taking the 10\,s ASKAP integration time as the uncertainty around the peak of AB1. 
    Using Equation \ref{eq:prob_coinc} with a temporal offset of $42$\,s, we obtain a probability of independent, coincident events of $3.2\times 10^{-8}$, strongly indicating a causal relation. Considering there were four additional lower-energy flares detected by TESS during ASKAP observations \citep{2019ApJ...884..160V}, and taking the total TESS flare rate of $1.49\,\text{days}^{-1}$ \citep{2019ApJ...884..160V}, we obtain a conservative trials-factor corrected coincidence probability of $7.8\times 10^{-6}$. This is an overestimation, because the rate of energetic flares such as the one detected on MBJD\,58605 is substantially lower than the total flare rate. This represents the first definitive association of an {interferometrically detected} coherent stellar burst with an optical flare in the literature.
    
    \subsection{Classification of Radio Bursts}
    The association of these radio bursts with the large optical flare indicates that they can be interpreted within the paradigm of solar radio bursts, which also are associated with multiwavelength activity. We interpret AB1 as a solar-like decimetric burst or an unresolved group of type III bursts. In the latter interpretation, we note that the measured frequency drift rate of $-4.7\pm 0.66$\,MHz\,s$^{-1}$ is not compatible with the positive drift rate of individual type III bursts, and instead may represent the `envelope' of the group of bursts. Solar observations show that both decimetric spike and type II bursts are often associated with hard X-ray bursts \citep{1986SoPh..104..145T, 1978SoPh...58..121S}. The occurrence of AB1 prior to the impulsive phase of the flare (see Figure \ref{fig:flare_timing}) suggests that particle acceleration was already ongoing during this early time \citep{1986SoPh..104..145T}. This is consistent with the findings of \citet{2006SoPh..236..369K}, who detected decimetric bursts similar to AB1 prior to the impulsive phase of a flare, which occurred high above the impulsive flare energy release site. They suggested that these decimetric bursts indicate a destabilization of the upper coronal magnetic field, which may be connected in some way to the flare energy release in the lower corona.
    
    The onset of AB2 is 35 minutes after the flare impulsive peak, and occurs during the gradual decay phase of the flare. This burst lasts for 119 minutes and has a complex morphology, exhibiting a variety of both positive and negative drift components. These details of AB2 are consistent with the $\sim 30$\,minute delays from solar flare peaks of the type IVdm bursts studied in \citet{2011ApJ...743..145C}, and of the complex early component of type IVdm bursts described in \citet{1967SoPh....1..304T}. AB2 is also elliptically polarized, a property also exhibited by some type IV bursts \citep{1963PASJ...15..195K}. 
    
    The radio light curves in Figure \ref{fig:multiwavelength}
    show that AB3 either directly follows, or temporally overlaps with AB2, and shares the same handedness of circular polarization. This suggests that they are two components of the same event, and are likely to originate from the same region in the stellar corona.
    The fractional bandwidth $\Delta \nu / \nu$ ranges from $0.1$--$0.3$, although the upper frequency cutoff is not within the ASKAP frequency range at the early stages of the burst, so the upper limit may be higher. The burst is up to 100\% circularly polarized, and exhibits a gradual drift of $-7.0\pm0.2$\,kHz\,s$^{-1}$ over its $353$\,minute duration. These properties are again consistent with solar type IV bursts, which display high degrees of circular polarization (up to 100\%), high brightness temperatures ($\sim 10^{11}$\,K; \citealt{2011ApJ...743..145C}), and can show relatively modest fractional bandwidths \citep{1976ApJ...204..597B}. The long-duration, slowly drifting morphology of the event also closely resembles other type IVdm events from the Sun (e.g., see Figure 5(a) from \citealp{2011ApJ...743..145C} or Figure \ref{fig:flare_spec}(d) from \citealp{1963PASJ...15..327T}). 
    
    The properties of AB2 and AB3 together, and their association with the 
    $1.6 \times 10^{32}$\,erg optical/H$\alpha$ flare, identify this radio event as a type IV burst. {Our detection of this event with full-Stokes dynamic spectroscopy along with the suite of multiwavelength observations make this the most compelling example of a solar-like radio burst from another star to date.}
    
    \subsection{Type IV Bursts as Indicators of Post-Eruptive Arcades and Coronal Mass Ejections}
    
    {As described in Section \ref{spaceweather_bursts}, studies of solar type IV bursts have established that these events are very closely associated with CMEs and SEPs \citep{1986SoPh..104...33R, 1988ApJ...325..895C, 1988ApJ...325..901C}, and indicate ongoing electron acceleration during magnetic field reconfiguration in the wake of CMEs \citep[e.g.][]{1992JGR....97.1619K}.
    Recently, a systematic study of solar type IV events by \citet{2020A&A...639A.102S} established that these bursts occur in columnar structures near the extremities of post-eruptive loop arcades, which they ascribe to the magnetic flux rope of the outgoing CME.} The association of post-eruptive loop arcades and CMEs with decimetric type IV bursts has also been noted by \citet{2011ApJ...743..145C} -- for example, the 2002 April 21 ``HF type IV'' burst presented in \citet{2004ApJ...600.1052K} and \citet{2011ApJ...743..145C} occurred at the same time as the rise of a post-eruptive loop system after an X1.5 flare \citep{2002SoPh..210..341G}. \citet{2011ApJ...743..145C} argued that these bursts may be originate in low-density flux tubes, with a field-aligned potential drop driving ECMI emission. Ultraviolet imaging of the rising post-eruptive loop arcades during the 2002 April 21 event showed the presence of evacuated flux tubes \citep{2002SoPh..210..341G}, supporting this claim, and perhaps explaining the existence of small degrees of linear polarization in other solar type IV bursts \citep{1963PASJ...15..195K, 1991A&A...249..250M}. 
    This picture, developed from multiwavelength observations of solar flares may explain the high degree of circular polarization, presence of linear polarization, and the long duration of the type IV burst from Proxima Cen. Driven by large-scale magnetic field reconfiguration associated with a post-eruptive loop arcade, the type IV burst from Proxima Cen is highly suggestive of a CME leaving the stellar corona \citep{1986SoPh..104...33R,1988ApJ...325..895C,2004A&A...422..337T}.
    
    If this type IV burst is indeed associated with a CME, it does not directly probe CME properties, but its exceptional nature indicates that eruptive processes on active M-dwarfs may only be associated with the most powerful flares. This is consistent with numerical simulations of CMEs and associated type II bursts by \citet{2020ApJ...895...47A}, {who reported that a simulated CME with a kinetic energy of $1.7\times 10^{32}$\,erg is weakly confined within a Proxima Cen-like magnetosphere. Applying a solar-like energy partition between bolometric flare energy and CME kinetic energy ($E_{K,\text{CME}}$) of $E_\text{bol}/E_{K,\text{CME}} = 0.3$ \citep{2012ApJ...759...71E}, we estimate a CME kinetic energy of $\sim 5.5\times 10^{32}\,\text{erg}$ for the putative CME indicated by our flare--type IV event. Even with a more conservative energy partition $E_\text{bol}/E_{K,\text{CME}} = 1$ \citep{2015ApJ...809...79O} (resulting in $E_{K,\text{CME}} = 1.6 \times 10^{32}\,\text{erg}$), these CME kinetic energy estimates closely match or exceed the weak CME confinement scenario explored by \citet{2020ApJ...895...47A}, depending on the flare-CME energy partition.}
    
    If energetic flares such as the $1.6 \times 10^{32}$\,erg flare presented here are necessary for eruptive space weather events, then their low rate ($\sim 0.04\,${day}$^{-1}$; \citealp{2018ApJ...860L..30H}) indicates that the space weather environment around Proxima Cen may be less threatening to planetary habitability than predicted by solar scaling relations \citep{2015ApJ...809...79O}. {However, without direct evidence for the putative CME and its influence on planetary companions, our inferences on the space weather environment around Proxima Centauri also remain indirect.}


    

\section{Conclusions}
\label{proxima_concl}
{We have presented simultaneous optical and radio observations of the nearby active M-dwarf and planet host, Proxima Cen. Photometric monitoring with TESS and the Zadko 1\,m Telescope reveal a large flare, with an estimated bolometric energy of $1.6 \times 10^{32}\,\text{erg}$. Simultaneous spectroscopic monitoring with WiFeS showed strong enhancements in H$\alpha$ and other chromospheric emission lines, indicating substantial deposition of energy into the chromosphere. Radio observations with ASKAP reveal a sequence of intense coherent bursts associated with this flare. The first, AB1, occurred $\sim 42\,\text{s}$ before the onset of the optical flare. The burst properties are consistent with solar-like decimetric spike bursts, which occur prior to the impulsive phase of flares, and indicate electron acceleration prior to the main impulsive outburst. The spectro-temporal and polarization properties of the radio bursts that trail the optical flare identify them as a type IV burst event. This is the most compelling identification of a solar-like radio burst from another star to date. Appealing to the properties of solar radio bursts, we suggest that the type IV burst from Proxima Cen is indicative of a CME, and ongoing electron acceleration in post-eruptive magnetic structures. 
Observational campaigns incorporating low-frequency radio, soft X-ray, and extreme ultraviolet observations may reveal in more detail the properties of post-eruptive loop systems on M-dwarfs, and directly probe CME properties. These will be required to verify the solar type IV burst-CME relationship on active M-dwarfs, and to probe the influence of these events on planetary companions. To this end, we strongly encourage further effort toward improving our understanding of space weather around active stars, {which remains limited in comparison to the solar case.}}

\section*{Acknowledgements}
We thank Gregg Hallinan and Stephen White for useful discussions, and Phil Edwards for providing comments on the manuscript. We thank the referee for their constructive feedback, which improved this work.
A.Z. thanks Thomas Nordlander and Gary Da Costa for their advice on observing with the ANU 2.3m Telescope. 
A.Z. is supported by an Australian Government Research Training Program Scholarship.
T.M. acknowledges the support of the Australian Research Council through Grant No. FT150100099.
D.K. is supported by NSF Grant No. AST-1816492.
Parts of this research were conducted by the Australian Research
Council Centre of Excellence for Gravitational Wave Discovery (OzGrav),
through Project No. CE170100004. M.A.M. acknowledges support from a National Science Foundation Astronomy and Astrophysics Postdoctoral Fellowship under Award No. AST-1701406.
Some of the data presented in this paper were obtained from the Mikulski Archive for Space Telescopes (MAST). STScI is operated by the Association of Universities for Research in Astronomy, Inc., under NASA contract NAS5-26555. Support for MAST for non-HST data is provided by the NASA Office of Space Science via grant NNX13AC07G and by other grants and contracts.
This research was supported by the Australian Research Council Centre of Excellence for All Sky Astrophysics in 3 Dimensions (ASTRO 3D), through Project No. CE170100013.
The International Centre for Radio Astronomy Research (ICRAR) is a Joint Venture of Curtin University and The University of Western Australia, funded  by the Western Australian State government.
ASKAP is part of the Australia Telescope National Facility, which is managed by the CSIRO. 
Operation of ASKAP is funded by the Australian Government with support from the National Collaborative Research Infrastructure Strategy. ASKAP uses the resources of the Pawsey Supercomputing Centre. Establishment of ASKAP, the Murchison Radio-astronomy Observatory, and the Pawsey Supercomputing Centre are initiatives of the Australian Government, with support from the Government of Western Australia and the Science and Industry Endowment Fund. 
We acknowledge the Wajarri Yamatji as the traditional owners of the Murchison Radio-astronomy Observatory site.

\vspace{5mm}
\facilities{ASKAP, ANU 2.3m Telescope (WiFeS), TESS, Zadko 1m Telescope.}

\software{\texttt{astroalign} \citep{2020A&C....3200384B}, Astropy \citep{2018AJ....156..123A}, \textsc{casa} \citep{casa}, ccdproc \citep{2015ascl.soft10007C}, matplotlib \citep{Hunter:2007}, NumPy \citep{van2011numpy}, SExtractor \citep{1996A&AS..117..393B}, PyWiFeS \citep{2014Ap&SS.349..617C}, \texttt{specutils}\footnote{\url{https://specutils.readthedocs.io/}}, emcee \citep{2013PASP..125..306F}.}

\bibliography{proxcenbib}{}
\bibliographystyle{aasjournal}

\end{document}